\documentstyle[12pt,prd,aps]{revtex}
\begin{document}
\baselineskip=17pt
\draft
\title{Cosmic Tau Neutrinos\footnote{Talk given at YITP Workshop on 
       Theoretical Problems related to Neutrino Oscillations, 28 February-01 
       March, 2000, Kyoto, Japan.}}
\author{Athar Husain\footnote{E-Mail: athar@phys.metro-u.ac.jp}}
\address{Department of Physics, Tokyo Metropolitan University, 
Minami-Osawa 1-1, Hachioji-Shi,\\ 
Tokyo 192-0397, Japan}

\maketitle
\begin{abstract} 
\tightenlines

The collisions between accelerated protons and the photons present in cosmos 
can be a main source of high-energy neutrinos ($E\, \geq 10^{6}$ GeV) of all 
flavors above the atmospheric neutrino background. I discuss here the 
possibility of photohadronic production of 
high-energy cosmic tau neutrinos in an astrophysical site and study some of 
the effects of vacuum neutrino flavor mixing on their subsequent propagation. 
I also discuss the prospects for observations of these high-energy cosmic tau 
neutrinos in new large neutrino telescopes.

\end{abstract}

\section{Introduction}

	High-energy  neutrino ($E\, \geq 10^{6}$ GeV) astroparticle physics is
now a rapidly developing field impelled by the need for improved flux 
estimates as well as a good understanding of detector capabilities for all 
neutrino flavors, particularly in light of recently growing experimental 
support for flavor oscillations \cite{skk}.

	Currently envisaged sources of high-energy cosmic neutrinos include,
for instance, cores of Active Galactic Nuclei (AGNs)\cite{prod,prod1}. 
Production of high-energy cosmic neutrinos other than the AGNs may also be 
possible \cite{review}.

	In near future, it may become feasible to distinguish between different
cosmic neutrino flavors using the information about their event topologies as
well as the  angle and energy resolutions from several of the large neutrino 
telescopes and/or (horizontal) shower arrays possibly with a combination of 
different detection techniques \cite{report}.

	This contribution is organized as follows: In section II, I briefly 
discuss the possibility of photohadronic production of cosmic tau neutrinos 
relative to electron/muon neutrinos, consider some of the effects of 
vacuum neutrino flavor mixing on their subsequent propagation and discuss some 
prospects for their observations. In section III, I summarize the discussion. 

\section{Cosmic Tau Neutrinos}

\subsection{Some flux estimates}

	I consider here briefly the photohadronic ($\gamma p$) production of 
high-energy cosmic
tau neutrinos in AGNs only. In $\gamma p$ collisions, the cosmic tau neutrinos
(and anti neutrinos) are mainly produced from the relevant decay chain of 
$D^{\pm}_{S}$ in $\gamma p \rightarrow D^{\pm}_{S}+X$.
Following the notation of Ref. \cite{berezin}, the cosmic tau neutrino flux
spectrum can be calculated in roughly three steps.  
The first step in estimating the cosmic tau neutrino flux spectrum is to 
calculate the yield of the relevant unstable hadron ($D^{\pm}_{S}$) in $\gamma
p$ collisions:
\begin{equation}
 Y_{D^{\pm}_{S}}=\int \int \int \mbox{d}E_{p}\mbox{d}E_{\gamma} 
 \mbox{d}\cos\theta
 (1-\cos\theta)\frac{n_{p}(E_{p})}{n_{p}(E_{D^{\pm}_{S}})}
 \frac{1}{\sigma}\frac{\mbox{d}\sigma}{\mbox{d}E_{D^{\pm}_{S}}}
 \frac{n_{\gamma}(E_{\gamma})}{n_{0}},
\end{equation}
where $\theta $ is the angle between photon and proton momentum in the 
laboratory system. The yield in Eq. (1) is essentially a result of folding the 
given photon spectrum, $n_{\gamma}(E_{\gamma})/n_{0}$ having energy 
$E_{\gamma}$ 
with the proton spectrum having a power law, $n_{p}(E_{p})\, \sim \, 
E_{p}^{-(\epsilon+1)} (
\epsilon\, \geq 1)$ while taking into account the details of relevant 
 kinematics. The differential cross-section can be taken from, 
for instance, from \cite{anjos}, whereas $n_{\gamma}(E_{\gamma})/n_{0}$ is 
taken from \cite{prod}. The second step is to convolve this yield 
with the relevant decay functions of $D^{\pm}_{S}$ to obtain the cosmic 
 tau neutrino
yield. This yield is further convolved with the proton spectrum to obtain the
cosmic tau neutrino flux. Finally, the cumulative cosmic tau neutrino flux 
spectrum is obtained 
by integrating the cosmic tau neutrino flux  over the luminosity function of 
the AGN while restoring the relevant numerical factors. A
notable characteristic feature in the shape of the (cumulative) neutrino  flux
 spectrum as a function
of neutrino energy is the flat region for relatively small neutrino energies
 because of kinematics and subsequently a steeply falling flux spectrum with 
 roughly the same 
spectral index as that for protons for relatively large neutrino energies, 
with rather smooth interpolation between these
two asymptotic behaviours. The principal difference between the tau and non 
tau (electron and muon) neutrino flux spectra is that of relatively lower 
value of the former as compared to latter with overall the same flux spectrum 
shape. The main reasons being: $\sigma_{\gamma p\rightarrow DX}/\sigma_{
\gamma p\rightarrow \pi X} \leq {\cal O} (10^{-3}-10^{-4})$ and secondly
BR($D \rightarrow \nu_{\tau})/
\mbox{BR}(\pi \rightarrow \nu_{\mu})\leq 3/99 \sim {\cal O}(10^{-2})$. 
Therefore, the total relative suppression is $\sim {\cal O}(10^{-4}-10^{-5})$ 
 essentially
for all relavant center of mass energies. It is relevant to mention that there
is no formation of resonance in $\gamma p \rightarrow D^{\pm}_{S}+X$ channel 
unlike in $\gamma p \rightarrow \pi^{\pm}+X$ channel for the relevant center 
of mass energies.

\subsection{Effects of vacuum neutrino flavor mixing}

It has been pointed out that there are no matter effects 
on vacuum neutrino flavor oscillations for relevant $\delta m^{2}$ values 
 [${\cal O} (10^{-10})\leq \delta m^{2}/\mbox{eV}^{2}\leq {\cal O} (1)$]
 for high-energy cosmic neutrinos originating from AGNs 
\cite{athar}. I, therefore consider here the effects of vacuum neutrino flavor 
mixing/oscillations only. 
For some other possible mixing effects, see \cite{npb}. 

The effects of vacuum neutrino flavor mixing in the context of three flavors 
are analytically discussed in \cite{9lomcon}. It was pointed out there that 
starting from the ratio of intrinsic cosmic neutrino fluxes as 
$F^{0}_{e} : F^{0}_{\mu} :F^{0}_{\tau} = 1 :2 : 0$\footnote{As discussed in
previous subsection, $F^{0}_{\tau}$ is negligibly small and therefore can be 
taken as 0.}, irrespective
of the flavor oscillation solution to the solar electron neutrino deficit, 
one obtains the ratio of final (downward going) cosmic neutrino fluxes as 
$F_{e} : F_{\mu} :F_{\tau} = 1 :1 : 1$. A somewhat detailed 
numerical study that properly incorporates the effects of non vanishing 
$\theta_{13}$ and the CP violating phase $\delta $ confirms that indeed this 
 is the case, i.e., the
normalized difference between any two neutrino flavors, 
$\Delta F_{\alpha \beta}\,  (\alpha \neq \beta; \alpha, \beta = e, \mu, \tau)$
 is typically of the order of 1\% \cite{japan}:

\begin{equation}
 \Delta F_{\alpha \beta} = \frac{|F_{\alpha}-F_{\beta}|}{\sum_{\alpha}
 F_{\alpha}}\leq {\cal O}(10^{-2}).
\end{equation}
An empirical determination of $\Delta F_{\alpha \beta}$, if become feasible,
may be useful to test the unconventional astrophysics/particle physics.

	Here, I consider the possibility that these three active light 
neutrinos mix with a fourth (sterile) flavor. 
This possibility accommodates the flavor oscillation solution for LSND anomaly
\cite{lsnd}
in addition to explaining the solar electron and atmospheric muon neutrino   
deficits also in terms of flavor oscillations with the introduction of a new 
$\delta m^{2}$ scale [$\delta m^{2}\sim {\cal O}(1)$ eV$^{2}$] relative to 
three flavor mixing scheme. If the effective number of light neutrino 
flavors from the Big-Bang nucleosynthesis is less than four then four 
($\theta_{13},\theta_{14},\theta_{23}, 
 \theta_{34}$) out of six mixing angles are small and can be
negelected \cite{kek,yasuda}. Following the description given in 
Ref. \cite{9lomcon}
and using the parameterization for the 4$\times $4 neutrino mixing matrix from 
Ref. \cite{kek}, one obtains the following $P$ matrix for vanishing $\delta$'s:

\begin{equation}
 P = \left( \begin{array}{cccc}
                             1 & 0   & 0   & 0 \\
                             0 & 1/2 & 1/2 & 0 \\
                             0 & 1/2 & 1/2 & 0 \\
                             0 & 0   &  0  & 1 \\
                            \end{array}
                      \right).
\end{equation}
Note that except the last row and column, this $P$ matrix is the same as that 
for SMA(MSW) situation in the context of three flavors. The absence of a 
relatively 
dense object is assumed here between the cosmologically distant AGNs 
($\sim $ 100 Mpc, where 1pc $\sim $ 3$\times 10^{18}$ cm) and the neutrino 
telescopes 
so as not to change this (vacuum) flavor oscillations pattern significantly. 
Also note that the neutrino energy dependence in relevant (vacuum) 
flavor oscillation probability expression is averaged out here for the entire
neutrino energy range relevant for observations.
 	Thus, starting from $F^{0}_{e} : F^{0}_{\mu} :F^{0}_{\tau}:F^{0}_{s} =
 1 :2 : 0: 0$, one obtains $F_{e} : F_{\mu} :F_{\tau}:F_{s} = 1 :1 : 1: 0$ at
the level of $F^{0}_{e}$. 
That is, in this (mass and) mixing scheme one also obtains a (downward going) 
cosmic 
tau neutrino flux comparable to corresponding cosmic non tau neutrino flux.

\subsection{Prospects for observations}

	Currently, two suggestions are available to identify the cosmic tau
neutrino flavor. Both suggestions are based on the relatively short decay 
lifetime of the $\tau$ lepton produced in deep inelastic scattering of cosmic
tau neutrinos with the nuclei.

	There is a suggestion of measuring cosmic tau neutrino flux through 
 double shower ({\it double bang}) events in underwater/ice  Cerenkov 
telescopes \cite{sandip}. A more recent suggestion is to detect a small 
pile up of upward going $\mu$-like events in the (10$^{4}$-10$^{5}$)~GeV 
energy range with a fairly flat zenith angle dependence \cite{halzen}.

	Using the flux estimates given in \cite{prod}, the cosmic tau 
neutrino induced downward going double shower event
rate for a typical km$^{2}$ surface area neutrino telescope in ice/water can 
be $\sim \cal{O}$ (1)/yr$\cdot$sr using Eq. (3) for $E\geq 10^{6}$ GeV.

	It may also become possible to obtain some useful information about 
the cosmic neutrino flavor from large (horizontal) shower arrays \cite{zas}.
For instance, the criterias of separableness and containedness of the two 
cosmic
tau neutrino induced showers as discussed in \cite{9lomcon,sandip} implies that
for the Pierre Auger array, the following essentially half an order of 
magnitude neutrino 
energy interval may be relevant: $5\times 10^{8}\leq E/\mbox{GeV}\leq 10^{9}$.
Note that the two showers develop here mainly in air.
However, the energy at which the two showers start separating in Pierre Auger 
lies below the planned threshold for detection in Pierre Auger so that
even in the case of the most favourable cosmic neutrino 
flux and oscillation probabilities, the expected number of events per year 
turns out to be  small.

\section{Conclusions}

 Irrespective of neutrino flavors, in each of the neutrino (mass and) mixing 
scheme discussed here, the final flux of downward going high-energy cosmic tau 
neutrinos is essentially comparable to that of non tau neutrinos, whereas 
intrinsically it is negligible. 

Prospective observation of cosmic tau neutrino flavor may become feasible in 
new large neutrino telescopes or/and in (horizontal) shower arrays.

\paragraph*{Acknowledgments.}

	This work is supported by a Japan Society for the Promotion
of Science fellowship. The author thanks O. Yasuda for comments.

\end{document}